\documentclass[preprint,authoryear,10pt]{elsarticle}

\usepackage{amssymb}
\usepackage{amsfonts}
\usepackage{amsmath}
\usepackage{amssymb}
\usepackage{subfig}
\usepackage{graphicx}

\biboptions{authoryear}

        \newcommand{\refequ}[1]{Eq.~(\ref{#1})}
        \newcommand{\paren}[1]{\left( #1 \right)}

\journal{Journal of the Mechanics and Physics of Solids}

\begin{document}

\begin{frontmatter}



\title{Interfacial separation between elastic solids with randomly rough surfaces: comparison between theory
and numerical techniques}


\author[lulea,juelich]{A.~Almqvist\corref{author}}
\author[ottawa]{C.~Campa\~{n}\'{a}}
\author[juelich,sumy]{N.~Prodanov}
\author[juelich]{B.N.J.~Persson}

\cortext[author] {\textit{E-mail address:} andreas.almqvist@ltu.se; Tel.: +49 920 492407; Fax: +46 920 491399}
\address[lulea]{Division of Machine Elements, Lule\aa~University of Technology, 971 87 Lule\aa, Sweden}
\address[juelich]{IFF, FZ-J\"ulich, 52425, J\"ulich, Germany}
\address[ottawa]{University of Ottawa, Department of Chemistry, Ottawa, K1N 6N5, Canada}
\address[sumy]{Sumy State University, 2 Rimskii-Korsakov Str., 40007 Sumy, Ukraine}

\begin{abstract}

We study the distribution of interfacial separations $P(u)$ at the contact region between
two elastic solids with randomly rough surfaces. An analytical expression is derived for $P(u)$
using Persson's theory of contact mechanics, and
is compared to numerical solutions obtained using (a) a half-space method based on the
Boussinesq equation, (b) a Green's function molecular dynamics technique and (c) smart-block
classical molecular dynamics. Overall, we find good agreement between all the different approaches.

\end{abstract}

\begin{keyword}
Contact mechanics, randomly rough surfaces, elastic solids, pressure distribution,
interfacial separation


\end{keyword}

\end{frontmatter}


%
\section{Introduction}
Modeling the contact mechanics between elastic solids with surfaces that are rough on multiple length
scales is a challenging task. To perform such a task several theoretical approaches have been
developed over the years. At the core of all the approaches lie approximations that relate to
describing the shape of the contacting surfaces. In a seminal paper, Greenwood and Williamson
\citep{greenwood1966} (GW) proposed that the contact problem between two elastic rough
surfaces could be reduced to the problem of one infinitely-hard rough surface acting on a
flat elastic counterface. Within their model, the rough topography was described by a large
collection of hemispherical asperities of uniform radius (which individually satisfied the
Hertzian approximation) with a height distribution that followed
a Gaussian law. This initial approach was later extended by Greenwood and Tripp \citep{greenwood1970}
by considering the presence of roughness on the two contacting surfaces. Further contributions to
the original GW methodology have been proposed by Whitehouse and Archard \citep{whitehouse1970}, Nayak
\citep{nayak1971}, Onions and Archard \citep{onions1973}, Bush et al. \citep{bush1975, bush1979}
and Whitehouse and Phillips \citep{whitehouse1978, whitehouse1982}. All these
models rely on the definition of ``asperity''. The asperity concept itself has proved quite controversial
and depends on the resolution of the instrument used to measure the surface profile\citep{poon1995}. Another
drawback of GW-type approaches is that using only a few parameters to describe the surfaces
generates a one-to-many mapping possibilities, i.e., the same set of parameters can be deduced for
surfaces obtained by completely different machining processes.

In spite of the great increase in computing power in the past
decade, analytical theories are still very much needed to understand the contact mechanics of
solids with surfaces that display roughness on more than three decades in length-scales.
Theoretical models to tackle such problems rely on
approximations and idealizations in order to analytically solve the equations of elasticity.
For example, such equations can be exactly solved for the contact problem of a parabolic tip acting
on a flat surface under the assumptions of linearly elastic, frictionless materials (Hertz model).
Similarly, an exact solution can be obtained for the contact between a sinusoidal elastic surface and a
rigid plane (Westergaard model) \citep{Westergaard_1939}. The GW model and its extensions are
further examples of how to deal with surface roughness in contact mechanics, resulting in simple
analytic formulas. These easy-to-handle formulas have proved to be of great importance, and are
frequently employed in the design process of new technical applications. The Hertz and the Westergaard
models provide accurate representation of the mechanics of single asperities. In addition, the GW 
model describes approximately the (low squeezing-pressure) contact between surfaces exhibiting
roughness on a single or a narrow distribution of length scales. However, most real surfaces have
roughness over many decades of length scales. Here, the long range elastic coupling between the
asperity contact regions, which is neglected in the GW model, is now known to strongly influence
contact mechanics \citep{elast,campanaJPCM08}. If an asperity is pushed downwards at a certain
location, the elastic deformation field extends a long distance away from the asperity influencing
the contact involving other asperities further away \citep{Bucher}. We note that the lateral coupling
between contact regions is important for arbitrary small squeezing pressure or load. The reason for
this is that surfaces with roughness on many length scales can be considered as consisting of large
asperities populated by smaller asperities, with the smaller asperities being populated by even
smaller asperities and so on. Thus, when two solids are squeezed together by a very small
external force, the distance between the macro-asperity contact regions will be very large,
and one may be tempted to neglect the elastic coupling between the macro-asperity contact regions.
However, the separation between the micro-asperity contact regions within a macro-asperity region
will in general be very small, and one cannot neglect the elastic coupling between such micro-asperity
contact regions. This latter effect is neglected in the GW theory, significantly limiting its
prediction capabilities when applied to most real surfaces. Additionally, in the GW model the asperity
contact regions are assumed to be circular (or elliptical) while the actual contact regions
(at high enough experimental resolution) show fractal-like boundary lines \citep{Borri,Pei,Chunyan1}.
Therefore, because of their complex geometries, one should try to avoid explicitly invoking the
nature of the contact regions when searching for an analytical methodology to solve the contact
problem of two elastic rough surfaces.

Recently, an analytical contact mechanics model that does not use the asperity concept
and becomes exact in the limit of complete contact has been developed by Persson
\citep{JCPpers,PSSR,PerssonPRL,Yang2008,Chunyan1}. The theory accounts for surface roughness on all
relevant length scales and includes (in an approximate way) the long range elastic coupling between
asperity contact regions. In this theory the information about the surface enters via the surface
roughness power spectrum $C({\bf q})$, which depends on all the surface roughness wavevectors ${\bf q}$
components. The theory can be used to calculate the interfacial stress distribution $P(\sigma,\zeta)$,
from which one can obtain the area of real contact as a function of the squeezing pressure $p$ and the
magnification $\zeta$. Furthermore, the theory predicts the average interfacial separation $\bar u$
for any applied external load.

Besides analytical approaches, numerical algorithms (deterministic) have also been developed to
understand the contact mechanics of elastic solids with rough boundaries. As the speed and memory
capacity of computers increase, numerical methods have become a viable alternative to analytical methods
when modeling surfaces of three-dimensional (3D) solids having surface roughness extending over at most three decades
in length-scales.
Nevertheless, simplifying assumptions about the material and the topography are still needed to
ensure reasonable computational time windows. Much is still to be done in order to reach the capacity
to numerically simulate real surfaces that may have roughness from the nanometer scale up to the
macroscopic size of the system which could be cm.

Numerous numerical works have been reported in the literature aiming to solve the contact mechanics of two linear elastic solids with rough surfaces. 
The majority of these are half-space models in which the elastic deformation is related to the stress field at the surfaces of the solids through integral equations where the domain of integration is the boundary of the half-space. This type of approach is commonly referred to as boundary element method (BEM). 
Twenty years ago Lubrecht and Ioannides \citep{lubrecht1991} suggested applying multilevel techniques to facilitate the numerical solution of the BEM. With the same objective in mind Ren and Lee \citep{ren1994} implemented a moving grid method to reduce storage of the influence matrix when the conventional matrix inversion approach is used to solve this type of problem.
Bj\"orklund and Andersson \citep{bjorklund_andersson1994} extended the conventional matrix inversion approach by incorporating friction induced deformations. Alternative techniques that aim to solve the elastic contact of rough surfaces are the Fast Fourier Transform (FFT)-based method introduced by Ju and Farris \citep{ju1996} and a follow-up extension, based on a variational principle \citep{kalker1977}, proposed by Stanley and Kato \citep{stanley1997}. The contact between solids with realistic surface
topographies under relatively small loads usually leads to plastic deformations. Tian and Bhushan, \citep{Tian1996} based their theoretical model is based on a variational principle for linear elastic perfectly plastic materials. In this way, not only the in-contact topography and the corresponding pressure distribution but also the unloaded plastically deformed topography can be obtained for the case when the loading is high enough
to cause yield. This model was further developed in the paper by \citep{sahlin2010a} and this is also the BEM employed in the present work, but here we restrict the analysis to linear elastic materials.

In earlier works, the prediction of Persson's contact mechanics theory for the interfacial stress
distribution $P(\sigma)$ and the contact area have been compared to numerical results obtained
using the finite element method (FEM) \citep{Hyun}, molecular dynamics \citep{Yang2008} and
Green's function molecular dynamics (GFMD) \citep{campana07epl}. In this paper, we will show how
the theory can be extended to also predict the distribution of interfacial separations $P(u)$.
This quantity is of crucial importance for problems like leak-rate of seals \citep{Chunyan1,Leak1,Leak2}
or mixed lubrication \citep{Pekl,mic,squeezeout}. The analytical results will be compared to numerical
solutions obtained using (a) a half-space method based on the Boussinesq equation (BEM), (b) a
Green's function molecular dynamics technique and (c) smart-block molecular dynamics.

%
\section{Contact mechanics theory of Persson}
\label{Persson}

Consider the frictionless contact between two elastic solids with Young's elastic moduli $E_0$ and
$E_1$ and Poisson ratios $\nu_0$ and $\nu_1$. Assume that the surfaces of the two solids
have height profiles $h_0 ({\bf x})$ and $h_1({\bf x})$, respectively. The elastic contact mechanics
for the solids can be mapped into that of a rigid substrate with height profile
$h({\bf x}) = h_0({\bf x})+ h_1({\bf x})$ and a second elastic solid with a flat surface and
Young's modulus $E$ and Poisson ratio $\nu$ chosen so that \citep{Johnson2}
\begin{equation}
\frac{1-\nu^2}{E} = \frac{1-\nu_0^2}{E_0}+\frac{1-\nu_1^2}{E_1}.
 \label{e50}
\end{equation}
The main physical variables that characterize the contact between the solids are the
stress probability distribution $P(\sigma)$ and the distribution of interfacial separations $P(u)$.
These functions are defined as follows:
$$P(\sigma ) = \langle \delta [\sigma - \sigma({\bf x})]\rangle, \ \ \ \ \ P(u ) = \langle \delta [u - u({\bf x})]\rangle$$ where $\delta(..)$ is the Dirac delta function, and $\sigma ({\bf x})$ and
$u({\bf x})$ are the stress and the interfacial separation at point ${\bf x} = (x,y)$, respectively.
The $\langle .. \rangle$ brackets denote ensemble averaging. Note that both $P(\sigma)$ and $P(u)$
have a delta function at the origin with its weight determined  by the area of real contact i.e. given by
$(1-A/A_0)\delta (\sigma)$ and $(A/A_0)\delta (u)$. Here $A_0$ is the nominal contact area
and $A$ the area of real contact projected on the $xy$-plane. Normalization conditions require that
$$\int d\sigma P(\sigma ) = 1, \ \ \ \ \ \int du P(u ) = 1$$
while
$$\int_{0^+}^\infty d\sigma P(\sigma ) = \frac{A}{A_0}, \ \ \ \ \ \int_{0^+}^\infty du P(u ) = \frac{A_0-A}{A_0}.$$
Thus from the interfacial distribution of stresses or separations one can immediately determine
the area of real contact $A$. The average interfacial stress (which must be equal to the applied
pressure) $\bar \sigma$, and the average interfacial separation $\bar u$, can be obtained as
$$\bar \sigma= \int d\sigma \ \sigma P(\sigma ), \ \ \ \ \ \bar u = \int du \ u P(u ).$$

The stress and interfacial separation distribution functions, $P(\sigma)$ and $P(u)$, are determined
by the elastic energy $U_{\rm el}$ stored in the asperity contact regions (see below).
The elastic energy $U_{\rm el}$ is written as \citep{BNJP,PSSR,elast}
\begin{equation}
 U_{\rm el} = \frac{E A_0}{4(1-\nu^2)} \int d^2q \ q C(q) W(q)
 \label{e32}
\end{equation}
where the surface roughness power spectrum is defined by
\begin{equation}
C(q) = \frac{1}{(2\pi)^2} \int d^2x \ \langle h({\bf x})h({\bf 0})\rangle {\rm e}^{-i{\bf q}\cdot {\bf x}}.
 \label{e53}
\end{equation}
The height profile $h({\bf x})$ of any rough surface can be measured routinely
nowadays on all relevant length scales using optical and stylus experiments.

For complete contact $W(q)=1$ rendering an exact result for the expression of the energy above.
In Ref.~\citep{BNJP} it was argued that $W(q)=P(q)=A(\zeta)/A_0$ is the relative contact area
when the interface is studied at the magnification $\zeta=q/q_0$ (where $q_0$ is the
small-wavevector cut-off,
usually chosen as $\pi /L$, where $L=\surd A_0$ is the linear size of the surface).
The qualitative explanation for such an argument is that the solids will mainly deform in the
regions where they make contact, and most of the elastic energy will arise from the contact regions.
Nevertheless, using $W(q)=P(q)$ assumes that the energy (per unit area) in the asperity contact
regions is just the average elastic energy (per unit area) as if complete contact would occur. This
does not take into account that the regions where no contact occurs are those regions where most
of elastic energy (per unit area) would be stored if complete contact would
occur. Hence, we expect smaller stored elastic energy (per unit area) in the asperity
contact regions than obtained using $W(q)=P(q)$. In Ref.~\citep{Yang2008,elast,Stiff} we found that
using
\begin{equation}
W(q)=P(q)\left [\gamma+(1-\gamma)P^2(q)\right ] = P(q)S(p,q),
 \label{e33}
\end{equation}
with $\gamma \approx 0.45$ gives good agreement between theory and numerical
calculations. Note that for complete contact $P(q)=1$ and hence $W(q)=1$ which reduces to
the exact result for the elastic energy in such a limit. On the contrary, in the limit of small
contact, $P(q)<< 1$ which yields $W(q) \approx \gamma P(q)$. For $\gamma \approx 0.45$ this results
in an elastic energy which is a factor of $0.45$ smaller than the elastic energy (per unit area)
stored in the contact region in the case of complete contact. Recently, in an independent study,
Akarapu et al. \citep{Akarapu} found a value of $\gamma=0.48$ after analyzing a variety of rough
surfaces in contact with roughness down to the atomic scale, variable Poisson ratio and
Hurst exponents of $H=0.5$ and $0.8$.

\begin{figure}
\includegraphics[width=0.35\textwidth,angle=0.0]{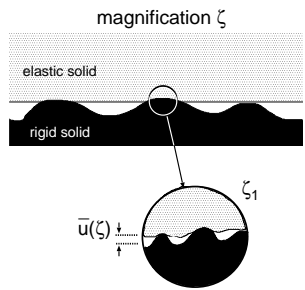}
\caption{\label{asperity.mag}
An asperity contact region observed at the magnification $\zeta$. It appears that
complete contact occurs in the asperity contact region, but when the magnification is
increased to the highest (atomic scale) magnification $\zeta_1$,
it is observed that the solids are actually separated by the average distance $\bar{u}(\zeta)$.
}
\end{figure}

The contact mechanics formalism developed by Persson \citep{JCPpers,PSSR,PerssonPRL,Yang2008} is
based on studying the interface between two contacting solids at different magnifications $\zeta$.
When the system is studied at the magnification $\zeta$ it appears as if the contact area
(projected on the $xy$-plane) equals $A(\zeta)$, but when the magnification
increases, it is observed that the contact is incomplete, and the surfaces in the apparent
contact area $A(\zeta)$ are in fact separated by the average distance $\bar u(\zeta)$,
see Fig. \ref{asperity.mag}. The (apparent) relative contact area $A(\zeta)/A_0$ at the magnification $\zeta$ is given by~\citep{JCPpers,Yang2008}
\begin{equation}
\frac{A(\zeta)}{A_0} = \frac{1}{(\pi G )^{1/2}}\int_0^{p_0} d\sigma \ {\rm e}^{-\sigma^2/4G}
= {\rm erf} \left ( \frac{p_0}{2 G^{1/2}} \right )
 \label{e51}
\end{equation}
where $p_0 = F_{\rm N}/A_0$ is the nominal squeezing pressure and 
\begin{equation}
G(\zeta) = \frac{\pi}{4}\left (\frac{E}{1-\nu^2}\right )^2 \int_{q_0}^{\zeta q_0} dq q^3 C(q) S(p,q).
 \label{e52}
\end{equation}
In most applications $A/A_0 << 1$ and in this case one may use $S\approx \gamma$.
The distribution of interfacial stress is given by (for $\sigma > 0$):
\begin{equation}
P(\sigma) = \frac{1}{2 (\pi G)^{1/2}} \left [ {\rm exp}\left (-\frac{(\sigma - p_0)^2}{4G}\right ) -
{\rm exp}\left (-\frac{(\sigma + p_0)^2}{4G}\right )\right ].
 \label{e52_}
\end{equation}

\begin{figure}
\includegraphics[width=0.45\textwidth,angle=0]{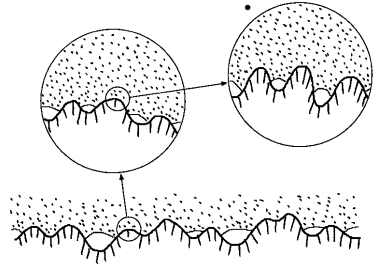}
\caption{
An elastic block (dotted area) in adhesive contact with a rigid rough substrate (dashed area).
The substrate has roughness on many different length scales and the block makes partial contact
with the substrate on all length scales. When a contact area is studied, at low magnification
it appears as if complete contact occurs, but when the magnification is increased it is observed
that in reality only partial contact has taken place.
}
\label{1x}
\end{figure}

Let us define $u_1(\zeta)$ to be the (average) height separating the surfaces which appear to come into
contact when the magnification decreases from $\zeta$ to $\zeta-\Delta \zeta$, where $\Delta \zeta$
is a small (infinitesimal) change in the magnification. $u_1(\zeta)$ is a monotonically decreasing
function of $\zeta$, and can be calculated from the average interfacial separation
$\bar u(\zeta)$ and $A(\zeta)$ using
(see Ref.~\citep{Yang2008})
\begin{equation}
u_1(\zeta)=\bar u(\zeta)+\bar u'(\zeta) A(\zeta)/A'(\zeta),
 \label{e54}
\end{equation}
where
\begin{equation}
\bar{u}(\zeta ) = \surd \pi \int_{\zeta q_0}^{q_1} dq \ q^2C(q) w(q) \int_{p(\zeta)}^\infty dp'
 \ \frac{1}{p'} S(p',q) e^{-[w(q,\zeta) p'/E^*]^2},
 \label{e55}
\end{equation}
$E^*=E/(1-\nu^2)$, $p(\zeta)=p_0A_0/A(\zeta)$
and
$$w(q,\zeta)=\left (\pi \int_{\zeta q_0}^q dq' \ q'^3 C(q') \right )^{-1/2}.$$

\begin{figure}
\includegraphics[width=0.45\textwidth,angle=0]{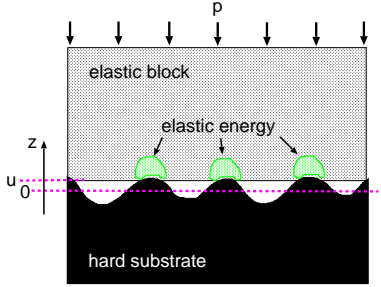}
\caption{\label{block}
An elastic block squeezed against a rigid rough substrate. The separation
between the average plane of the substrate and the average plane of the lower
surface of the block is denoted by $u$. Elastic energy is stored in the block in the vicinity
of the asperity contact regions.
}
\end{figure}

The definition of the distribution of interfacial separations $P(u)=\langle \delta [u-u({\bf x})]\rangle$
involves an ensemble average over many realizations of the surface roughness profile. If the surface
roughness power spectra has a roll-off wavevector $q_{\rm c}$ which is much larger than $q_0=\pi /L$,
where $L$ is the linear size of the surface, then performing an ensemble average is identical to
averaging over the surface area. In this case we can write the distribution of interfacial separations
as
\begin{equation}
  P(u) = \frac{1}{A_0} \int_{A_0} d^2x \ \delta(u-u({\bf x})).
  \label{e:Bo1}
\end{equation}
The probability distribution is normalized
\begin{equation}
  \int du \ P(u) = 1.
  \label{e:Bo77}
\end{equation}
In the contact mechanics theory of Persson \citep{Yang2008} the interface is studied at different
magnification $\zeta$. As the magnification increases, new short length scale roughness can be detected,
and the area of (apparent) contact $A(\zeta)$ therefore decreases with increasing magnification. The
(average) separation between the surfaces in the surface area which (appears) to move out of contact
as the magnification increases from $\zeta$ to $\zeta+d\zeta$, is denoted by $u_1(\zeta)$ and is
predicted by the Persson theory (see above). The contact mechanics theory of Persson does not directly
predict $P(u)$ but rather the probability distribution of separation $u_1$ (see Ref. \citep{Yang2008}):
\begin{equation}
  P_1(u) = \frac{1}{A_0} \int_1^{\zeta_1} d\zeta [-A'(\zeta)]\ \delta(u-u_1(\zeta)).
  \label{e:Bo3}
\end{equation}
Since $u_1(\zeta)$ is already an average, the distribution function $P_1(u)$ will be more narrow
than $P(u)$, but the first moment of both distributions coincide and is equal to the average
surface separation:
$$\bar u = \int_0^\infty du \ u P(u) = \int_0^\infty du \ u P_1(u). $$
To derive an approximate expression for $P(u)$ we write \refequ{e:Bo1} as
\begin{equation}
  P(u) = \frac{1}{A_0} \int_1^{\zeta_1} d\zeta [-A'(\zeta)]\ \langle \delta(u-u({\bf x}))\rangle_\zeta.
  \label{e:Bo4}
\end{equation}
Here $\langle ..\rangle_\zeta$ stands for averaging over the surface area which moves out of contact
as the magnification increases from $\zeta$ to $\zeta + d\zeta$. Note that
\begin{equation}
  \langle u({\bf x})\rangle_\zeta = u_1(\zeta).
  \label{e:Bo5}
\end{equation}
A surface which moves out of contact as the magnification increases from $\zeta$ to $\zeta+d\zeta$
will have short-wavelength roughness with wavevectors larger than $q>\zeta q_0$. Thus the separation
between these surface areas will not be exactly $u_1(\zeta)$, but will fluctuate around this value.
One may take this into account by using
\begin{equation}
  \langle (u({\bf x})-u_1(\zeta))^2\rangle_\zeta \approx h^2_{\rm rms}(\zeta),
  \label{e:Bo6}
\end{equation}
where $h^2_{\rm rms}(\zeta)$ is the mean of the square of the surface roughness amplitude including
only roughness components with the wavevector $q>q_0\zeta$. We can write
\begin{equation}
  h^2_{\rm rms}(\zeta) = \int_{q> q_0\zeta} d^2q \ C(q),
  \label{e:Bo7}
\end{equation}
where the surface roughness power spectra $C(q)$ can be calculated from the measured surface
topography. Using the definition
$$\delta (u) = \frac{1}{2 \pi}\int d\alpha \ e^{i\alpha u},$$
one can rewrite \refequ{e:Bo4} as
$$P = \frac{1}{A_0} \int d\zeta \ [-A'(\zeta)] \frac{1}{2 \pi} \int d\alpha \ \langle e^{i\alpha (u-u({\bf x}))}\rangle_\zeta$$
$$ = \frac{1}{A_0} \int d\zeta \ [-A'(\zeta)] \frac{1}{2 \pi} \int d\alpha \ e^{i\alpha (u-u_1(\zeta))} \langle e^{i\alpha (u_1(\zeta))-u({\bf x}))}\rangle_\zeta . $$
To second order in the cummulant expansion
$$P \approx \frac{1}{A_0} \int d\zeta \ [-A'(\zeta)] \frac{1}{2 \pi} \int d\alpha \ e^{i\alpha (u-u_1(\zeta)) -\alpha^2 \langle (u_1(\zeta)-u({\bf x}))^2\rangle_\zeta /2}, $$
or using \refequ{e:Bo6}:
$$P \approx \frac{1}{A_0} \int d\zeta \ [-A'(\zeta)] \frac{1}{ \left (2\pi h^2_{\rm rms}(\zeta)\right )^{1/2}} {\rm exp} \left ( - \frac{(u-u_1(\zeta))^2}{2 h^2_{\rm rms}(\zeta)}\right ) . $$
The above expression does not satisfy the normalization condition \refequ{e:Bo77}. We will therefore
use instead
\begin{eqnarray}
P \approx \frac{1}{A_0} \int d\zeta \ [-A'(\zeta)]
\frac{1}{\left ( 2\pi h^2_{\rm rms}(\zeta) \right )^{1/2}}
\nonumber\\
\times \left [ {\rm exp} \left (
    -\frac{(u-u_1(\zeta))^2}{2 h^2_{\rm rms}(\zeta)}\right ) +
    {\rm exp} \left ( - \frac{(u+u_1(\zeta))^2}{2 h^2_{\rm rms}(\zeta)}\right )\right ] .
  \label{e:Bo8}
\end{eqnarray}
The added term in this expression can be considered as resulting from the cummulant expansion of
$$\frac{1}{A_0} \int_{A_0} d^2x \ \delta(u+u({\bf x})) . $$
Note that such a term vanishes for $u>0$.

The theory described above predicts that for small squeezing pressures $p$ the area of real
contact is proportional to the squeezing pressure, while the interfacial separation depends
logarithmically on $p$.
Both results are related to the fact that when increasing $p$ existing contact areas grow and
new contact areas are formed in such a way that, in the thermodynamic limit (infinitely-large system),
the interfacial stress distribution, and also the size distribution of contact spots, are
independent of the squeezing pressure as long as these distributions are normalized to
the real contact area $A$~\citep{P3}.
In Ref.~\citep{LP} (see also \citep{LCS}) experimental results were presented to test the
dependence of $\bar u$ on $p$. In the study a rubber block was squeezed against an asphalt road
surface, and good agreement was found between the theory and experiments. The fact that $A\sim p$ for
small load is also well tested experimentally, and is usually considered as the explanation for
Coulomb's friction law which states that the friction force is proportional to the load or normal
force.

\begin{figure}[hbtp]
\centering
\begin{tabular}[b]{ll}
\includegraphics[width=0.4\textwidth]{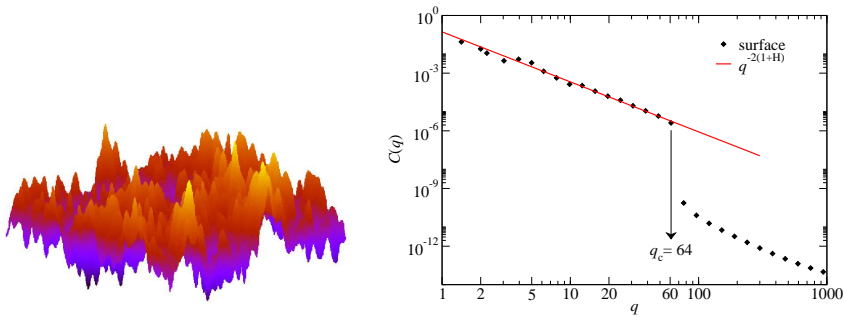} & \includegraphics[width=0.5\textwidth]{Fig3b.ps} \\
\end{tabular}
\caption{Graphical representation of a rough topography with Hurst exponent $H=0.3$
and its corresponding height-height correlation function
$C(q)=\langle \vert h ({\bf q}) \vert^2 \rangle$ in Fourier
space. The surface topography was created using a Fourier
filtering technique and a hard cutoff $q_c=64$ was imposed to it in Fourier space
(in units of $2\pi/L$,where $L$ is the linear size
of the simulation cell). The continuous line represents the ideal algebraic scaling
expected for the height-height correlation function of such a surface.}
\label{surf_corr}
\end{figure}

\section{Numerical methods}
\label{numeric}

When two elastic solids with rough surfaces come into contact, the elastic deformations
perpendicular to the contacting plane extend into the solids a characteristic length $\lambda$ that
could be as large as the contacting plane's lateral size $L$. Thus, in order to properly
capture the mechanical response of the solids within the contact region, the elastic properties
of the material have to be considered up to a distance $L$ in the normal direction to the
contacting plane. This is why standard algorithms that use a full representation of the system
display a computational effort that scales with the system's size $L^3$. Such a scaling rapidly
becomes a limitation when the linear dimension of the solids increases.
The previous arguments explain why coarse-grained
numerical techniques are needed when studying the contact mechanics of solids with more than
two decades in surface roughness length-scales.

The theory developed in Sec. \ref{Persson} for the distribution of interfacial separations
$P(u)$ will be compared to the predictions of the following three different coarse-grained numerical
methods: (a) a half-space method based on the Boussinesq equation (BEM), (b) a Green's
function molecular dynamics technique (GFMD) and (c) smart-block classical molecular dynamics (MD).
For this, we have considered the contact between an elastic block with a flat
bottom surface and a randomly rough rigid substrate. Self affine fractal topographies with Hurst
exponent values of $H=0.3$, $0.5$ and $0.8$ (corresponding to the fractal dimension
$D_{\rm f} = 3-H = 2.7$, $2.5$ and $2.2$) are used to model the rigid substrate.
Fig.~\ref{surf_corr} shows the surface topography $h({\bf x})$ and the power spectrum $C({\bf q})$
(on a log-log scale) of one of our surfaces with $H=0.3$, hard cutoff $q_c=64$ (in units of
$2\pi/L$, where $L$ is the linear size of the simulation cell) and rms-slope $\sim 0.03$.


Randomly rough substrate profiles were generated on a two-dimensional square grid with
$2048\times2048$ mesh points. The surface heights were obtained via a Fourier Filtering Algorithm.
In the case of the GFMD and BEM methods the elastic interactions within the original elastic block
are chosen such that both Lame coefficients satisfy $\lambda=\mu=1$. This choice of the coefficients
results in a Young modulus of $E = 5/2$, a bulk modulus of $K = 5/3$, and a Poisson ratio of
$\nu = 1/4$.

In the smart-block molecular dynamics simulations we used a smaller system size than in the
other two numerical schemes. The surfaces were obtained by choosing every
4'th grid point from the original surfaces with $2048\times2048$ resolution. This procedure yielded
surfaces with $N_x \times N_y=512\times 512$ mesh points. Since the original surfaces were quite 
smooth at short length scales (see the power spectrum in Fig.~\ref{surf_corr}) the surfaces used 
in the smart-block MD simulations are expected to give the same
contact mechanics as the original topographies. This was confirmed by comparing the MD results
obtained for the $512 \times 512$ system sizes to those of the BEM method for the 
equivalent $2048 \times 2048$ systems.
In our MD simulations the atoms in the bottom layer of the
block are located on a simple square lattice with lattice constant $a = 2.6$~\AA. The lateral
dimensions of the block and substrate are $L_{x} = L_{y} = N_{x}a = 1331.2$~\AA.
The Young modulus of the block is $E = 250$~GPa and its Poisson ratio $\nu = 1/4$ identical to that
of the other two schemes.
Since no natural length scale exists in elastic continuum mechanics, one can directly compare the
results of the smart-block MD model to those of the methods (a) and (b) by simply using a distance
scaling factor of $(512/2048) a = 0.65$~\AA, and a pressure (or stress) scaling
factor of $250/2.5 = 100 \ {\rm GPa}$.

Next, we will provide a short technical review of the three numerical methods that we have employed 
in this work.

\subsection{Boundary Element Method}
\label{bem}
Any contact mechanics problem can be solved by using a technique that minimizes the total potential energy of the system. Assuming frictionless linear elastic contact, the variational problem including constraints to be solved can be expressed as \refequ{e:variational} (see, e.g.,
\citep{kalker1977}, \citep{Tian1996} and \citep{sahlin2010a}):
\begin{equation}
         \underset{p \geq 0}{\min }\paren{ \frac{1}{2}
           \int_{A} d^2x \ p ({\bf x}) u_z({\bf x}) -
           \int_{A} d^2x \ p ({\bf x}) u_z^*({\bf x})
         }, \label{e:variational}
\end{equation}
where $p({\bf x})$ is the pressure distribution,
$u_z({\bf x})$ is the associated elastic deformation and
$u_z^*({\bf x})$ is the prescribed surface displacements, equivalent to the roughness height coordinate ($h$) plus a constant controlling the prescribed shift.
The first term describes the internal complementary energy due to elastic deflection, and the second term governs the contribution
from the prescribed displacement $u_z^*({\bf x})$. Note that $u({\bf x})=u_z({\bf x})-u_z^*({\bf x})$.
The Boussinesq relation between pressure and elastic deformation employed for this work may be formulated as
\begin{equation}
         u_z({\bf x})=\frac{1-\nu^2}{\pi E}\int d^2x'
           \frac{p({\bf x'})}{|{\bf x}-{\bf x}'|},
         \label{e:boussinesq}
\end{equation}
For the BEM method employed here, the complete system of equations consists of \refequ{e:variational} and
\refequ{e:boussinesq}, and the
force balance relation
\begin{equation}
         \int_{A} d^2x \ p({\bf x}) = F_{\rm N}, \;\;\; p({\bf x}) \geq 0,
         \label{e:forcebalance}
\end{equation}
where $F_{\rm N}$ is the normal force or the applied load.

Numerically, the solution is achieved by employing the method outlined by \citep{sahlin2010a}, where the FFT algorithm is
utilized to accelerate the computation of the elastic deflection. For the numerical results presented in this work,
convergence of the solution process is reached when the following two convergence criteria are met:
\begin{itemize}

\item[-]The force balance criterion that controls that the load generated by the contact pressure supports the applied load.
$$\frac{1}{F_{\rm N}}\left| F_{\rm N}-\int_{A} d^2x \ p({\bf x}) \right|< 10^{-3},$$

\item[-]A geometric criterion that controls that points `in-contact' lie sufficiently close to the
contact plane.
$$\frac{\underset{{\bf x}\in A}{\max}\left|u_z({\bf x})-u_z^*({\bf x})\right|}{
\underset{{\bf x}\in A_0}{\max}\; h({\bf x}) - \underset{{\bf x}\in A_0}{\min} \;h({\bf x})} < 10^{-5}.$$
\end{itemize}
%
%
%

\subsection{Greens Function Molecular Dynamics}
\label{gfmd}

Greens Function Molecular Dynamics (GFMD)~\citep{campana06} is among the many simulation
techniques available to find the equilibrium configuration of a mechanical system under
the action of external loads. To achieve its goal GFMD solves the system's equations of
motion for a set of initial/boundary conditions.

The potential energy of a linear elastic (harmonic approximation) solid is given by
\begin{equation}
V = \frac{1}{2}  \sum_{ij}\sum_{\alpha \beta} k_{ij}^{\alpha \beta} u_{i\alpha} u_{j\beta} =
\frac{1}{2} \sum_{ij}  {\bf u}_i \cdot K_{ij} \cdot {\bf u}_j
\end{equation}
where ${\bf u}_i =\sum_\alpha u_{i\alpha} {\bf e}_\alpha$ (where ${\bf e}_1 =\hat x$, ${\bf e}_2 =\hat y$ and ${\bf e}_3 =\hat z$ are orthogonal unit base vectors)
is the deformation field at locations ${\bf x}_i$,
and $K_{ij}=\sum_{\alpha \beta} k_{ij}^{\alpha \beta} {\bf e}_\alpha {\bf e}_\beta$ the force constant matrix. In thermal equilibrium,
the ${\bf u}_i$ displacements will comply with the Boltzmann statistics with second
moments given by
\begin{eqnarray}
\langle {\bf u}_i {\bf u}_j \rangle & = &
\frac{1}{Z}
\int d{\bf u}_1 \cdots d{\bf u}_N \; {\bf u}_i \, {\bf u}_j \; e^{-\beta V}
\nonumber\\
& = &
k_B T \left[ K^{-1} \right]_{ij},
\label{eq:fluc_rel}
\end{eqnarray}
with $Z$ being the partition function. The equation above shows that
one can obtain the force constant matrix $[ K^{-1}]_{ij}$ from
measuring the fluctuations in the second moments $\langle {\bf u}_i {\bf u}_j \rangle$.
Furthermore, if the bottom surface of the solid is exposed to an external force
field ${\bf F}_{\rm ext}(\{ {\bf u}\})$, all the
degrees of freedom ${\bf u}_i$ that do not belong to the bottom surface can be integrated out
(eliminated) yielding an equivalent problem for the bottom surface
\begin{equation}
V = \sum^{surface}_{ij} \frac{1}{2} {\bf u}_i \cdot \tilde{K}_{ij} \cdot {\bf u}_j -
\sum^{surface}_i {\bf F}_{\rm ext}^{(i)}\cdot {\bf u}_i
\label{eq:real_space_V}
\end{equation}
where $\tilde{K}_{ij}$ are new renormalized force constants. Renormalization takes place such
that the new 2D surface obtained will deform under the action of the external field in
exactly the same way as the bottom surface of our original 3D-solid. The matrix
$$\left[ G \right]_{ij}=\left[ \tilde{K}^{-1} \right]_{ij}=\frac{\langle {\bf u}_i {\bf u}_j \rangle}{k_B T}$$
is known as the Greens' function of the system.
From \refequ{eq:real_space_V} one obtains the equilibrium condition
\begin{equation}
\sum_j \tilde{K}_{ij} \cdot {\bf u}_j = {\bf F}_{\rm ext}^{(i)}.
\label{eq:mosion}
\end{equation}
If the system is periodic in the
$(x,y)$-plane (translational symmetry) one can use the Fourier
transform to obtain decoupling of the modes in the ${\bf q}=(q_x,q_y)$-space as
\begin{equation}
\tilde {{\bf u}}({\bf q}) = \tilde{G}({\bf q})\tilde{\bf F}_{\rm ext}({\bf q}).
\label{eq:u_q}
\end{equation}

\noindent \refequ{eq:u_q} eliminates the non-local nature of the real-space
solution of \refequ{eq:mosion} while rendering an easily parallelizable scheme that
can be used to simulate the contact mechanics of large systems.
For a given interaction kernel $\tilde{G}({\bf q})$, it is
the implementation of Eq.~(\ref{eq:u_q}) in a molecular dynamics fashion
that lies at the core of GFMD.

Our GFMD implementation followed the approach described in previous
works~\citep{campana07epl,campanaJPCM08} where all the roughness is placed on the rigid substrate
and the elasticity on a flat GFMD block. The interactions between the block and the rough
substrate are modeled via a hard-wall potential. If the $z$-coordinate of
an atom within the elastic block at location ${\bf x}=(x,y)$ crosses through $h({\bf x})$, the
corresponding interaction energy increases from zero to infinity. To obtain the value of the surface
height $h({\bf x})$ inside any square element of the grid, we employed interpolation via bi-cubic
splines with zero partial and cross-derivatives at the corner grid points of each element.

Defining which block's atoms are in contact after equilibrium has been
reached is done by analyzing the pressure distribution.
In a fully equilibrated simulation, a wide pressure gap of several
orders of magnitude would exist between the atoms that belong to the
contact region and those which do not. This approach to defining the
contacting status of a certain atom differs from geometrical ones
where contact is defined by comparing the relative distance between
the block's atom and the corresponding surface height at the local
atomic position, as it is done within the BEM method.


\subsection{Smart-block Molecular Dynamics}\label{MDsetup}

The smart-block molecular dynamics (MD) system is composed of an elastic block interacting with a rigid
randomly rough substrate. The substrate and the bottom layer of the block consist of an array of
$512 \times 512$ atoms. Periodic boundary conditions are applied in the $xy$-plane. The atoms in the
bottom layer of the block form a simple square lattice with lattice constant $a = 2.6$~\AA. The mass
of a block atom is 197~amu, and its elastic parameters have already been mentioned at the
beginning of this section.

In order to allow for a correct description of the long-wavelength components in the deformation field of the block, its thickness
is chosen to be 1350.7~\AA, which is slightly larger than its lateral dimension. The technical details of
the smart-block implementation has been discussed elsewhere, see \citep{Yang2006}.
The current smart-block consists of 12 atomic layers, and merging factors of 2 (in all 3 directions) are
used for all layers, except the 1'st, 6'th and the 11'th. The smart-block contains 615780 atoms, and the total
number of atoms involved in the simulations is 877924.

The atoms at the block-substrate interface interact via a repulsive potential
$U(r)= 4\varepsilon\left(r_{0}/r\right)^{12}$, where $r$ is
the interatomic distance and the parameter $\varepsilon$ corresponds to the binding energy between
two atoms at the separation $r = 2^{1/6}r_{0}$.
In our calculations we have used the values $r_{0} = 3.28$~\AA~and $\varepsilon = 18.6$~meV.
Zero temperature is maintained during the simulations using a Langevin thermostat \citep{Griebel2007} and
the equations of motion have been integrated using Verlet's method \citep{Griebel2007,Rapaport2004} with
a time step of $\Delta t = 1$~fs.

In the present study the squeezing process proceeds as follows. The upper surface of the smart-block
is moved towards the substrate at a constant velocity of $v = 5$~m/s with the
block being compressed as its
bottom layer approaches the substrate. The duration of simulations depends on the type of
substrate and they last until a small enough separation between the bottom block layer and the substrate
is achieved. Note that the thickness of the smart-block may influence the results. In particular,
a too thin smart-block leads to noisy data with considerable deviation from the results of the other
methods. Increasing the thickness of the smart-block beyond the lateral size of the surface ensures
convergence in the results. One must add that the time dependencies of $p$ and $u$ are not
monotonic, and some oscillations are observed, which may be attributed to elastic waves propagating
in the block during its compression. Lowering the velocity of movement down to 2.5~m/s leads to
a decrease in the amplitude of these oscillations. Moreover, the large $u$ region in the dependence of pressure on $u$ gets closer to the BEM results when a lower value of $v$ is employed. This suggests that smaller values of $v$ should be used in future studies.
%
\section{Results and discussion}

\begin{figure}
[!htb]
\begin{center}
\includegraphics[
width = 0.7\textwidth
]{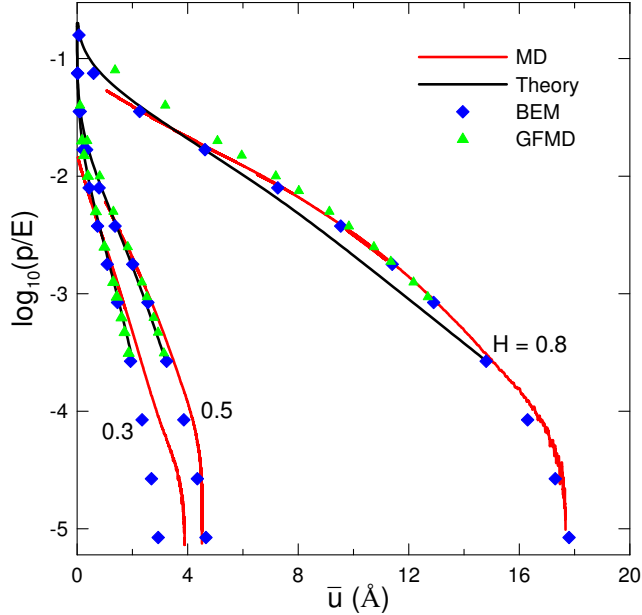}
\caption{The relation between the applied squeezing pressure $p$ and the average interfacial separation $\bar u$.}
\label{f:p_vs_sep}
\end{center}
\end{figure}

We first consider the dependency $p(\bar u)$ of the pressure $p$ on the average interfacial separation $\bar u$.
Figure~\ref{f:p_vs_sep} shows $p(\bar u)$ obtained using the analytical theory (black lines), the BEM (blue symbols),
GFMD (green symbols) and the smart-block MD approach (red lines).
The figure includes results for surfaces with Hurst exponents of
$H=0.3$, $0.5$ and $0.8$. Note that the analytical theory has been developed for infinite
systems, which will have infinitely high asperities, therefore always leading to a certain degree of contact
between the solids. The small systems sizes utilized in the numerical simulations
resulted in the highest asperities exceeding by only a factor of three (above the average plane) the
root-mean-square roughness of the surfaces. This explains the sharp drop in the pressure 
in the computer simulation curves at the height threshold value established by the tallest asperity.

\begin{figure}[!htb]
  \centerline{\includegraphics[width=0.7\textwidth]{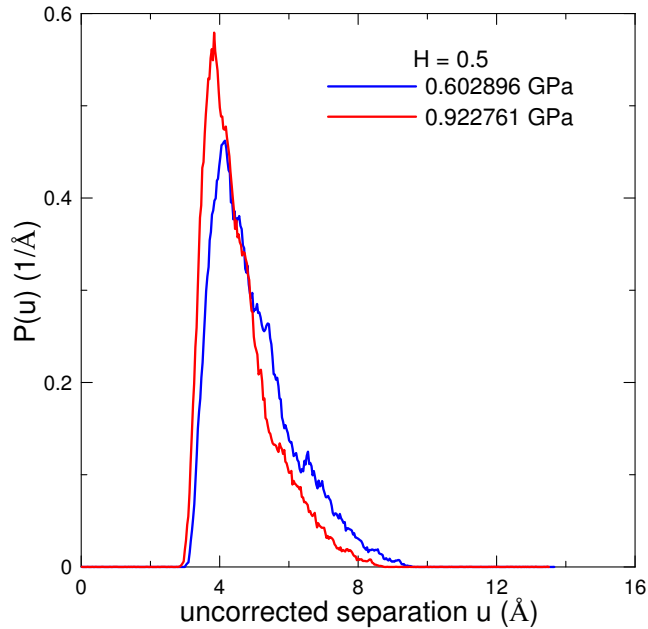}}
  \caption{Distribution of interfacial separations for the substrate with $H = 0.5$ obtained in
classical MD simulations. }
  \label{distribution_md}
\end{figure}

In the smart-block MD simulations, the finite range of the interaction potential between
the block atoms and the surface atoms resulted in a non-unique way of defining $\bar u$.
In the current work, to obtain $\bar u$ a small contribution $\delta u$ of about $4 \ {\rm \AA}$ 
has been subtracted from the difference $z_1-z_0$ in the average position of the interfacial atoms of the block and the substrate.
How to find the best $\delta u$ is a rather difficult question, and several alternative ways exist 
to accomplish such a goal. Here we have used that in the continuum limit the $P(u)$ distribution
must display a maximum (delta-like behaviour) at its origin. In smart-block simulations the maximum
in $P(u)$ gets shifted to a non-zero surface separation due to the finite range of the
interaction potential. This behavior is shown in Fig.~\ref{distribution_md} for two different external
pressure values. Thus, shifting the MD probability distribution $P(u)$ towards the origin by $\delta u$ is 
necessary in order to be able to compare with the GFMD and BEM continuum mechanics results.

\begin{figure}[hp]
        \centering
                \subfloat[$p/E=0.0032.$]
                {\includegraphics[width=0.31\textwidth]
                {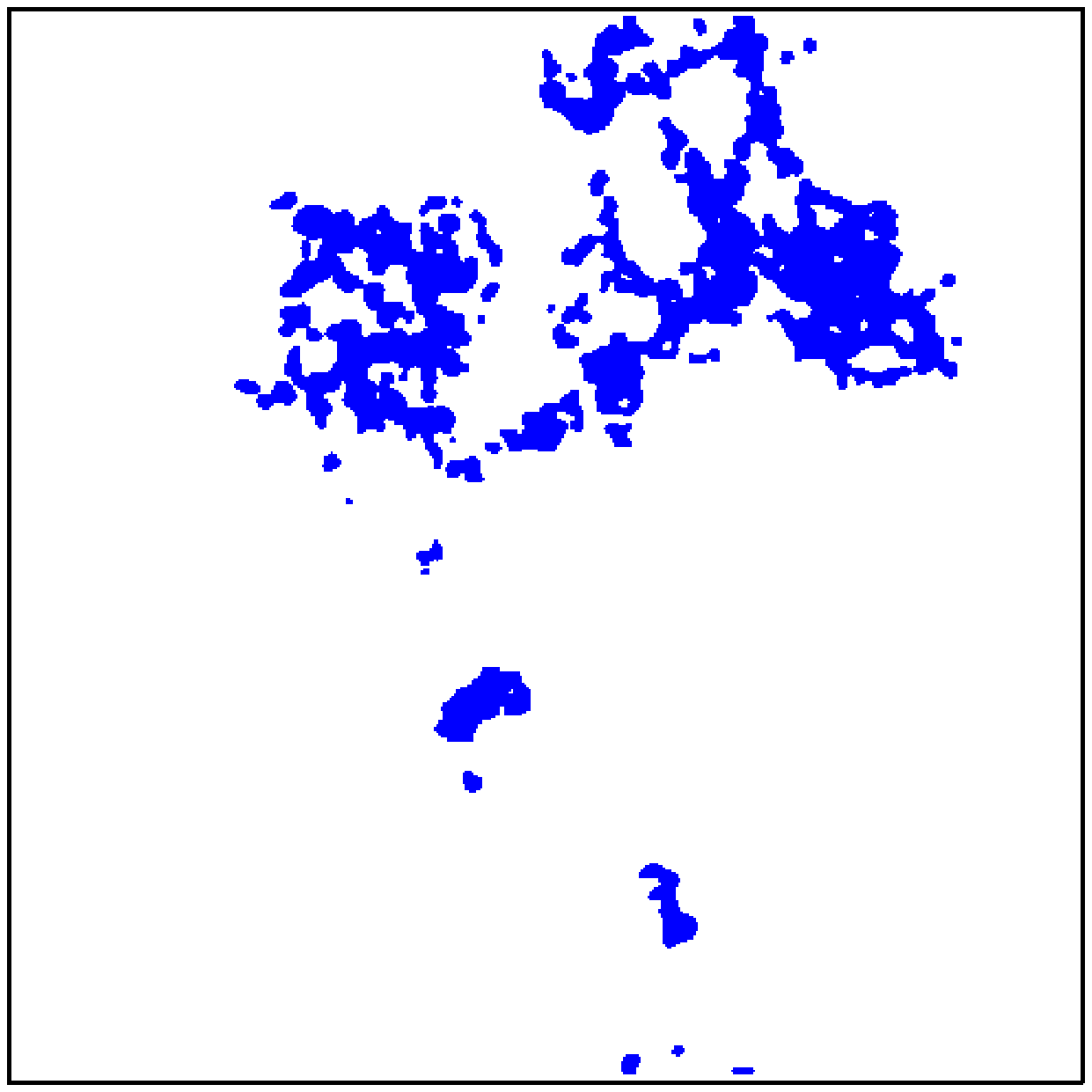}}\label{f:morph-a}
                \subfloat[$p/E=0.0080$.]
                {\includegraphics[width=0.31\textwidth]
                {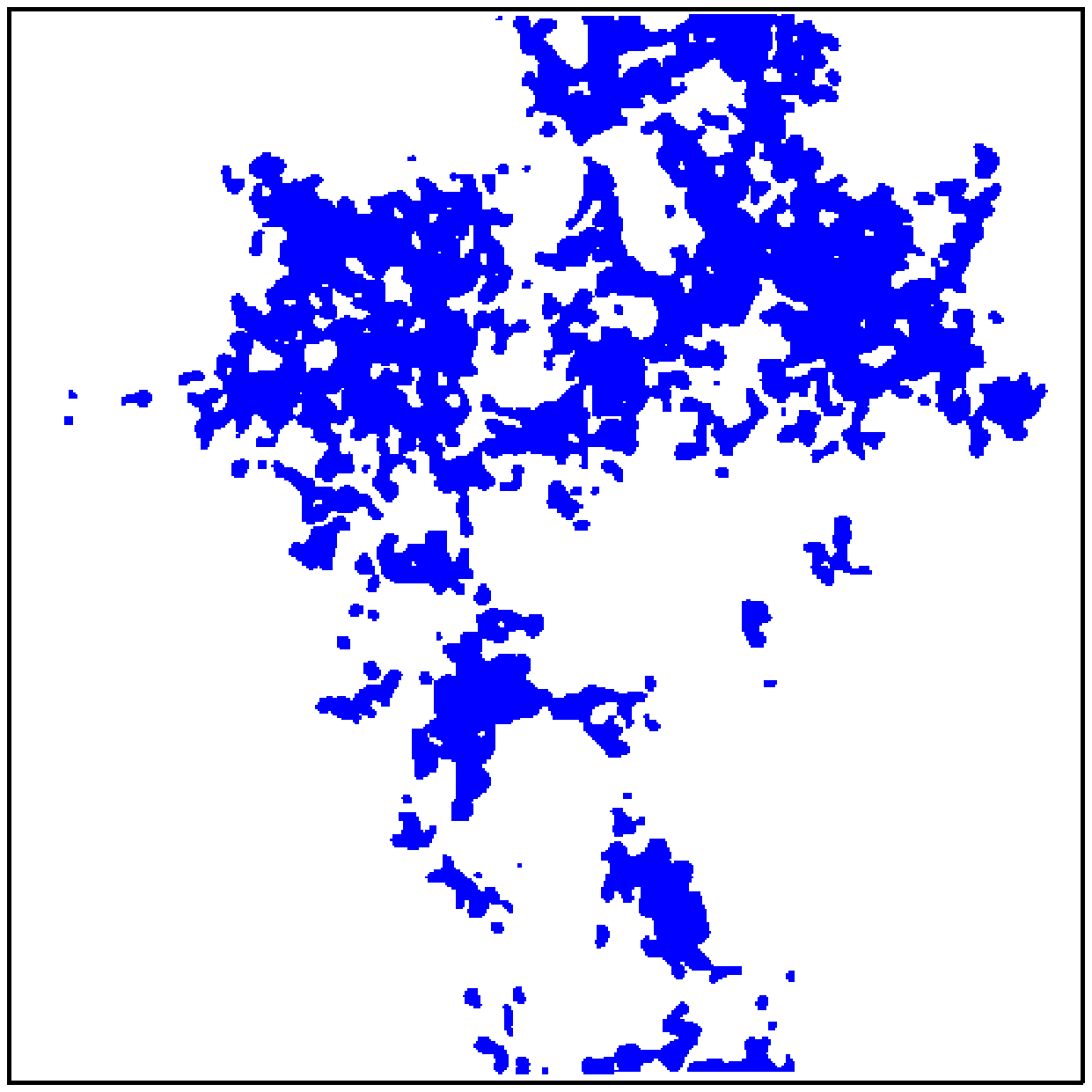}}\label{f:morph-b}
                \subfloat[$p/E=0.0120$.]
                {\includegraphics[width=0.31\textwidth]
                {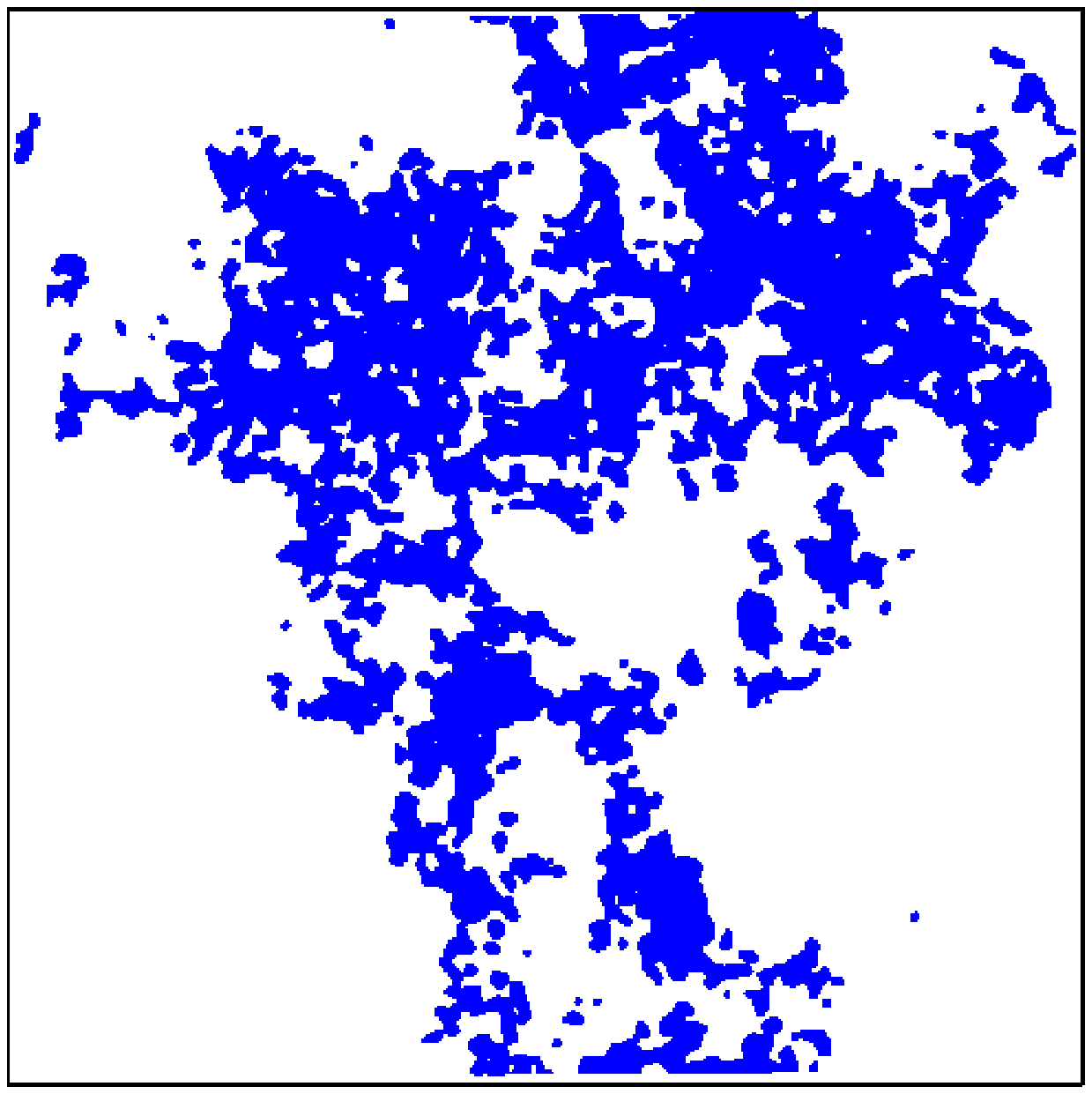}}\label{f:morph-c}
                \caption{Contact morphologies for the $H=0.8$ surface at three different
                contact pressures; (a) $p=0.0032E$, (b) $p=0.0080E$ and (c) $p=0.0120E$ when
                long-range elastic deformations are considered. The fractional contact areas are
                $A/A_0 = 0.066$, $0.161$ and  $0.238$, respectively}
        \label{f:contact_morphology}
\end{figure}

\begin{figure}[hp]
        \centering
                \subfloat[$A/A_0 = 0.066$.]
                {\includegraphics[width=0.31\textwidth]
                {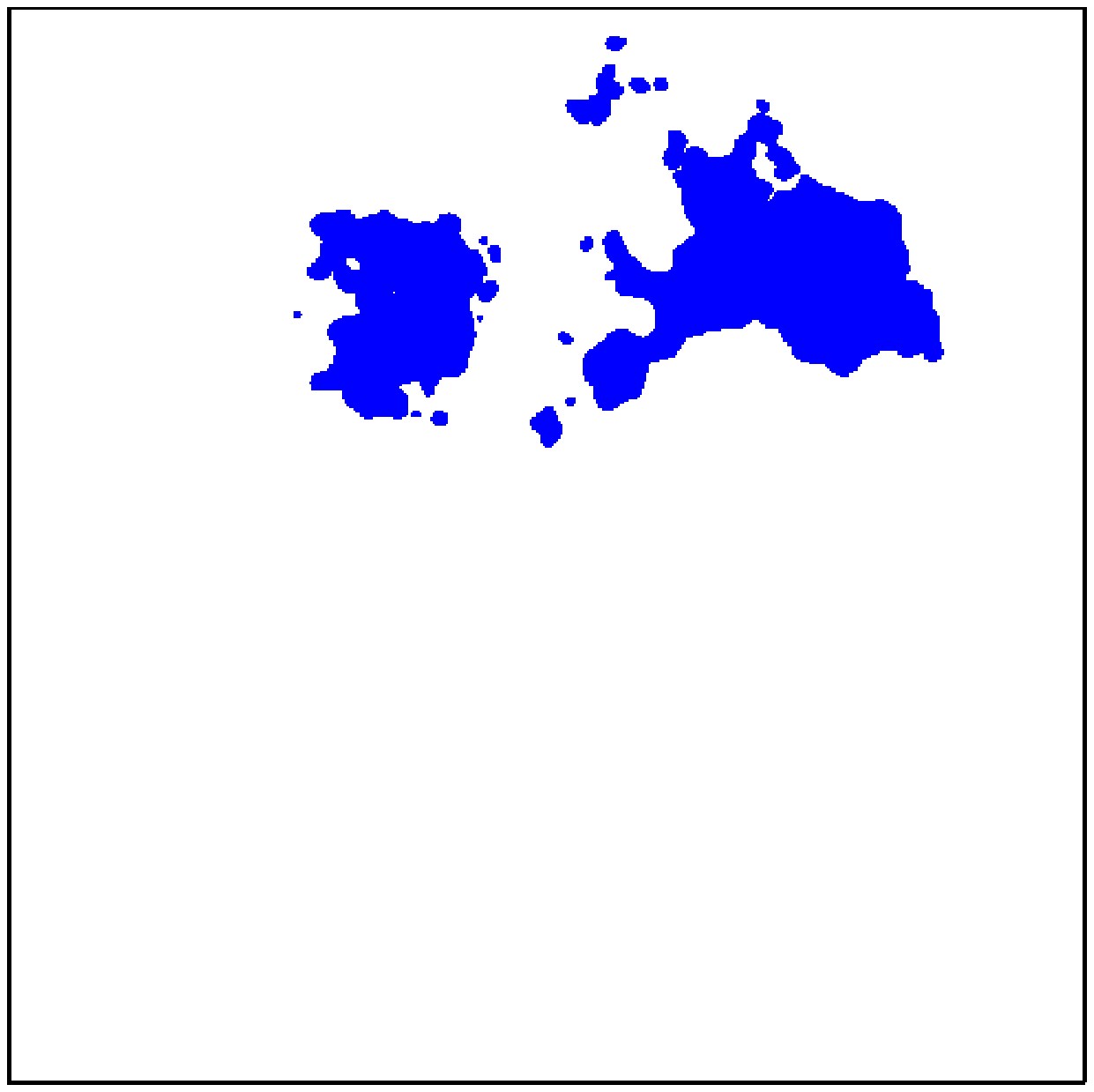}}\label{f:morphcp-a}
                \subfloat[$A/A_0 = 0.161$.]
                {\includegraphics[width=0.31\textwidth]
                {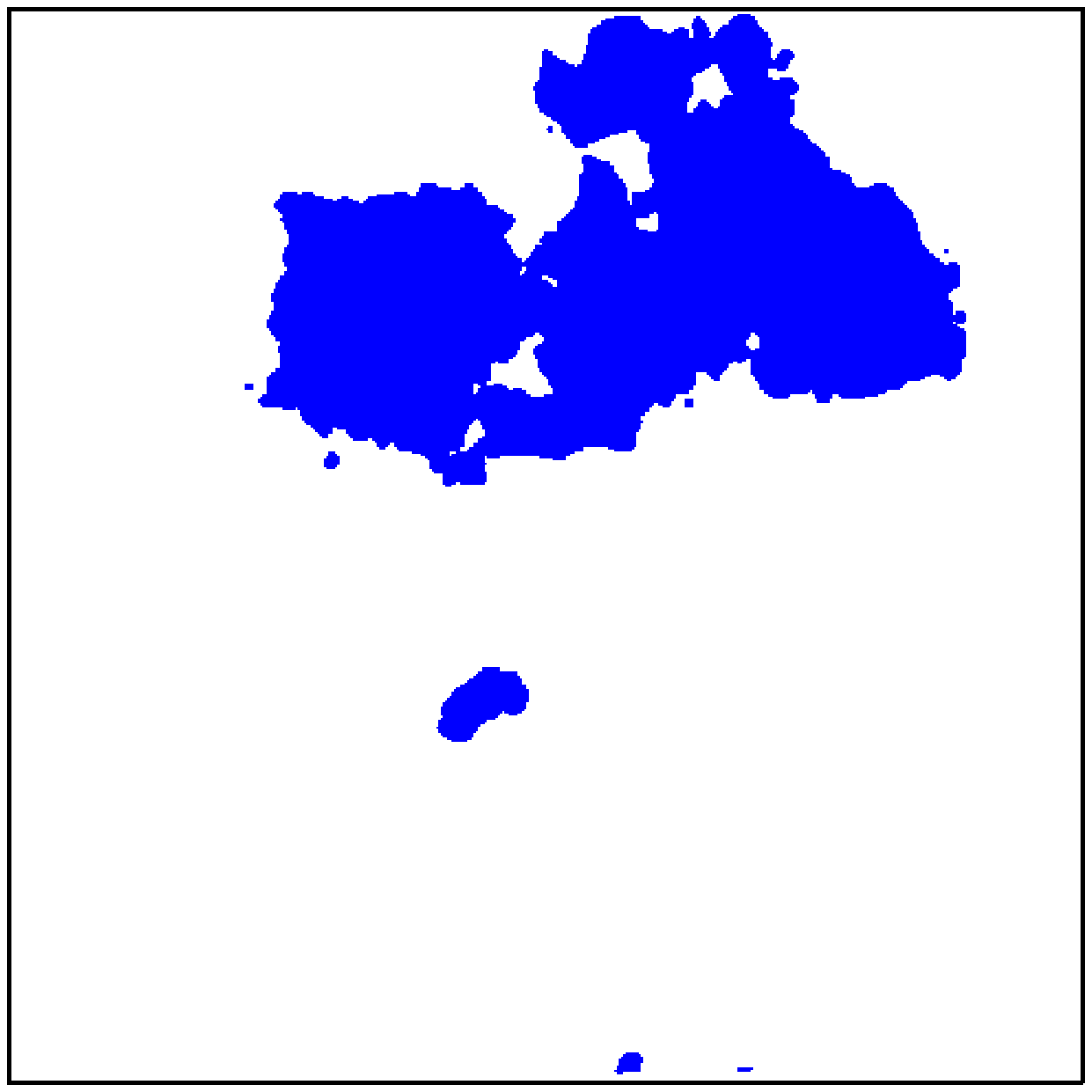}}\label{f:morphcp-b}
                \subfloat[$A/A_0 = 0.238$.]
                {\includegraphics[width=0.31\textwidth]
                {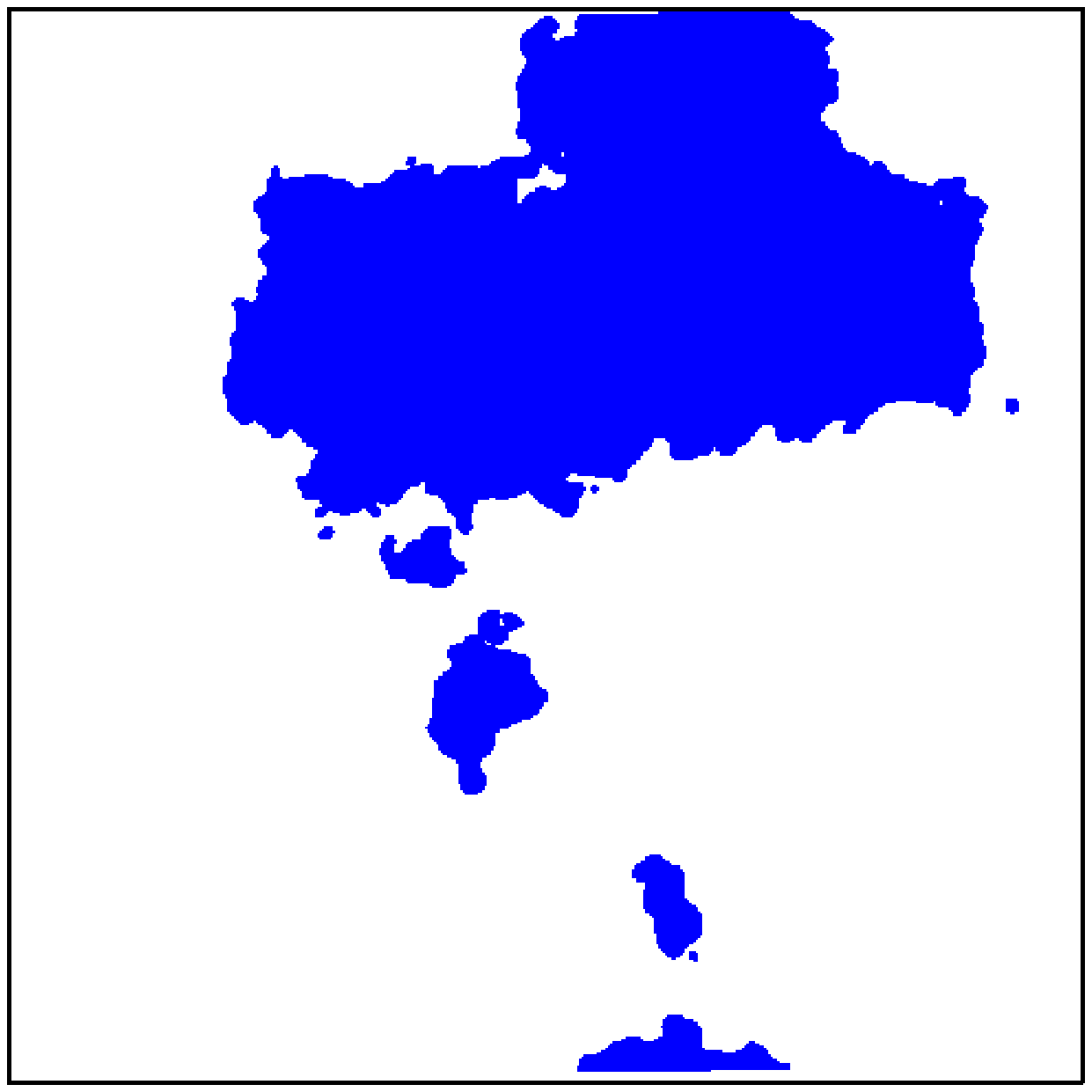}}\label{f:morphcp-c}
                \caption{Contact morphologies predicted by a bearing area model for the same
                ``true'' contact area $A$ values as in Figure~\ref{f:contact_morphology} but where
                long-range elastic coupling has been neglected.}
        \label{f:contact_morphology_cut_plane}
\end{figure}

Persson's contact mechanics theory predicts that the average interfacial separation
in the large $\bar u$ range is related to the applied pressure $p$ via $p\sim {\rm exp}(-\bar u /u_0)$,
where $u_0$ is of order of the root-mean-square roughness of the undeformed rough surface.
This result, already confirmed by experimental works~\citep{LP,LCS}, differs drastically
from the prediction of GW-like asperity models which instead yield a dependency of the type
$p \sim {\rm exp}(-b \bar u^2)$ with $b$ being a constant. The origin for those differences,
an exponential decay predicted by Persson's theory and a Gaussian decay obtained
by GW, is the omission of the long-range elastic deformations within asperity contact models
which also results in very different morphologies for the contact regions as illustrated in
Figs.~\ref{f:contact_morphology} and \ref{f:contact_morphology_cut_plane}.

Figure~\ref{f:contact_morphology} shows the contact morphology generated for a $H=0.8$ rough
surface at three different external pressures, $p=0.0032 E,\ 0.008 E,\ 0.012 E$, when the long-range
elastic deformations are taken into account. In Figure~\ref{f:contact_morphology_cut_plane}
we show the morphologies obtained (for identical values of the ``true" contact area $A$ ) when a
bearing area model is utilized. The lack of long-range elastic coupling in the latter model produces
qualitatively different contact morphologies (compare Figures \ref{f:contact_morphology} and \ref{f:contact_morphology_cut_plane}). 
Thus, when elastic deformations are considered, the contact regions become less compacted, with fractal-like boundaries, and are
distributed over a larger fraction of the nominal
contact area than those predicted by the asperity model in Fig.~\ref{f:contact_morphology_cut_plane}.

%

\begin{figure}[ptb]
\begin{center}
  \includegraphics[width = 0.6\textwidth]{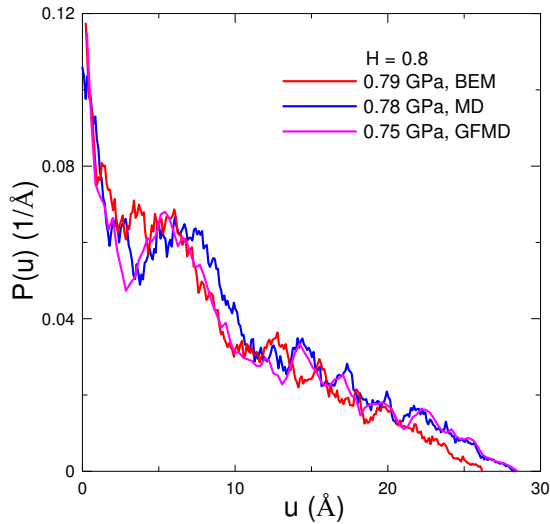}
  \caption{Probability distribution $P(u)$ of interfacial separations $u$, obtained using the BEM,
MD and GFMD methods for $p=0.79$, $0.78$ and  $0.75 \ {\rm GPa}$, respectively.}
  \label{f:distribution_H08_3methods}
\end{center}
\end{figure}

\begin{figure}
[ptb]
\begin{center}
\includegraphics[width=1.0\textwidth,angle=0.0]{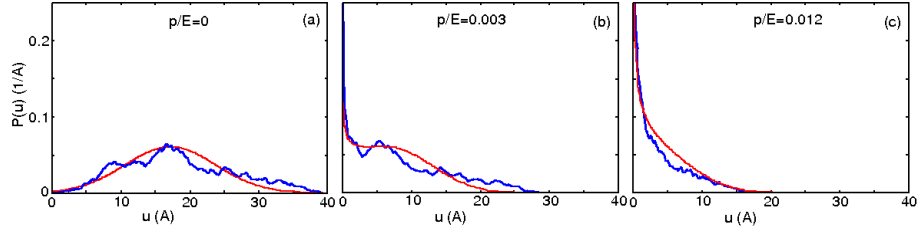}
\caption{Probability distribution $P(u)$ for several squeezing pressure values.
The red line corresponds to the theoretical predictions and the blue to numerical simulations (in (a)
and (c) using BEM and in (b) using GFMD).
}
\label{1u.2ProbabilityInterfacialSeparation}
\end{center}
\end{figure}

A first comparison between the probability distribution $P(u)$ of interfacial
separations $u$ obtained using the BEM, smart-block MD and GFMD methodologies at the squeezing pressures
$p=0.79$, $0.78$ and  $0.75 \ {\rm GPa}$, respectively, is displayed in Figure~\ref{f:distribution_H08_3methods}.
As shown, slight changes in applied pressure did not vary significantly the general shape of $P(u)$.
This is indicative of the individual consistency in the implementation of each numerical
technique. Further comparison of the numerical methods to the theoretical predictions is
included in Fig.~\ref{1u.2ProbabilityInterfacialSeparation} in the limit of (a) non-contact, (b) low-pressure
region and (c) medium-to-high pressure region. The red lines correspond to the predictions of Persson's
contact mechanics theory (\refequ{e:Bo8}) while the blue lines have been obtained from GFMD (in (b)) and BEM ((a) and (c)).
All the numerical results correspond to a single realization of the rough surface which explains the
rather large fluctuations (noise) in the $P(u)$ data. In particular, the ensemble averaged $P(u)$ for
the non-contact case (zero squeezing pressure, $p=0$) must follow a Gaussian law as given by the theory curve.
However, the lack of a small wavevector roll-off (or cutoff) in the surface roughness power
spectra implied that numerous independent topographies would have to 
be considered in order to achieve averages that closely represent the thermodynamic limit. 
Due to the large computational effort involved in such a task no
further attempt to improve our statistics was performed.

\begin{figure}
[ptb]
\begin{center}
\includegraphics[width=0.60\textwidth,angle=0.0]{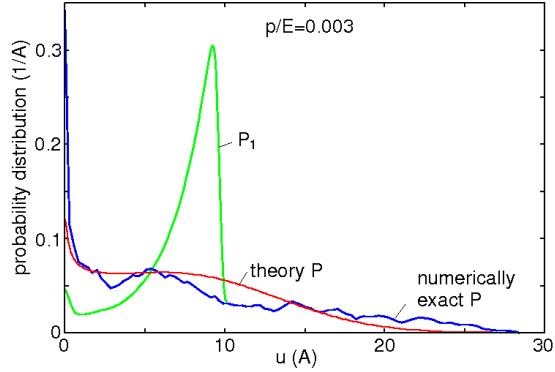}
\caption{Probability distribution $P(u)$ for the squeezing pressure $p=0.003 E$ obtained in a numerically
exact calculation (blue curve), and by applying the theoretical result from (\refequ{e:Bo7}) (red curve).
Also shown is the theory prediction (\refequ{e:Bo8}) for the distribution $P_1(u)$ of boundary-line
averaged interfacial separations (green curve).
}
\label{1u.2P}
\end{center}
\end{figure}

As already mentioned in the methods section, the original contact mechanics theory of Persson does not
directly predict $P(u)$ but instead $P_1(u)$ (distribution of boundary-line averaged interfacial separations)
which is a much sharper function than the interfacial separation distribution. The individual behaviour
of both functions for the $H=0.8$ surface at a squeezing pressure of $p=0.003 E$ is plotted in
Fig.~\ref{1u.2P}. In the figure the analytical $P(u)$ predicted
using \refequ{e:Bo8} (red curve) and  $P_1(u)$ (green line) (\refequ{e:Bo3}) are compared to the numerically exact
distribution generated from GFMD simulations (blue curve). Because the bin size of the $P(u)$
and $P_1(u)$ computations was smaller than that of the numerical GFMD study the delta function
at the origin $u=0$ barely shows in the first two cases. Nevertheless,
based on the results presented in Figs.~\ref{f:distribution_H08_3methods},
~\ref{1u.2ProbabilityInterfacialSeparation} and \ref{1u.2P}  we feel confident to conclude
that the extension presented in this work to compute  $P(u)$ within the framework of Persson's
original contact theory yields quite reasonable quantitative predictions when
compared to numerically exact simulations of the same contact problem.


\begin{figure}[!htb]
\begin{center}
  \includegraphics[width = 0.7\textwidth]{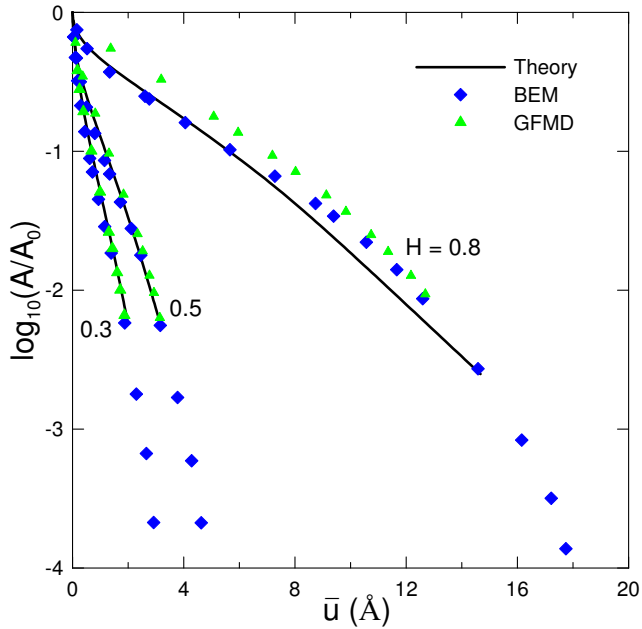}
  \caption{The contact area $A/A_0$ as a function of average interfacial separation $\bar u$ for the $H=0.3$, $0.5$ and $0.8$ surfaces. }
\label{f:ar_vs_sep}
\end{center}
\end{figure}

Another physical variable of interest in contact mechanics studies is the ratio
$A/A_0$ between the real area of contact and the nominal contact area. With the help of Persson's
theory one can derive expressions that relate $A/A_0$ to the average
interfacial separation $\bar u$. Next, the predictions from such a relation can be compared to those
of numerical calculations. Figure~\ref{f:ar_vs_sep} depicts the fractional contact ratio
obtained in theoretical and numerical simulations of rough surfaces with variable Hurst
exponents over a wide pressure region. As depicted, in the low-pressure
regime (large $\bar u$ zone) all approaches converged to the same limit.
This is further proof of the suitability of the numerical techniques discussed here to study
the contact mechanics of elastic solids with randomly-rough surfaces under engineering conditions.
As the pressure and $H$ exponent increase, discrepancies arised between GFMD and the other approaches.
In future works the cause for such discrepancies in the high pressure region must be further investigated.

%

%
%
\section{Summary and conclusions}

The contact mechanics theory developed by Persson has been extended
to allow for the calculation of the distribution of interfacial separations $P(u)$. The theory has
been applied to study the contact mechanics of a flat elastic solid squeezed against an infinitely-hard
randomly rough substrate. Three different coarse-grained numerical approaches have been used
to simulate the same problem: (a) the boundary element method
(BEM), (b) Green's function molecular dynamics (GFMD) and (c) smart-block molecular dynamics. The theoretical
predictions have been compared to those of the numerical methods.

All the numerical methods and the analytical theory gives very similar results
for the pressure $p$ and the fractional contact area $A/A_0$, as a function of the
average interfacial separation $\bar u$. In agreement with some earlier
numerical studies when find a linear proportionality between real area of contact
$A$ and the applied load at low loads, and a logarithmic relation between the average interfacial
separation and the applied pressure in the same load regime~\citep{Akarapu,Stiff,LP,LCS}.
We note that these functional
forms are those predicted by Persson's theory thus pointing to the capabilities of the theory to
properly account for long-range elastic deformations.

The distribution of interfacial separations $P(u)$ was studied in the low-to-medium pressure regions
and our results showed that the theory gives quantitative predictions of reasonable
accuracy for $P(u)$ when compared to the results obtained from numerically-exact calculations.
However, the rather small system sizes used in the numerical calculations resulted in finite size
effects (noise) for large interfacial separations. Among the numerical schemes we note that the BEM model,
based on a FFT-accelerated implementation of the Boussinesq equation, is fast
and accurate over the whole range of squeezing pressures. The GFMD method is also
computationally very fast, but its results deviated from the expected solution for large values
of the roughness exponent and high pressures. While the cause for such discrepancies needs to
be investigated, we note that it lies within the current
numerical implementation of the GFMD code and not in the GFMD theory which is in principle exact.
Lastly, the smart-block classical MD is
the most computationally demanding approach. Nevertheless, in contrast to BEM and GFMD,
it naturally includes adhesion and friction in an atomistic way, and its current
implementation can be applied within the full pressure range.

%

\vskip 0.5cm
{\bf Acknowledgments}

This work, as part of the European Science Foundation EUROCORES Program FANAS, was supported from funds
by the DFG and the EC Sixth Framework Program, under contract N ERAS-CT-2003-980409. A.~Almqvist acknowledges the EC Seventh Framework Program under contract ALM-PERG06-GA-2009-250623 and the Swedish Research Council VR under contract 621-2008-3839 for financial support.


\begin{thebibliography}{00}


\bibitem[Akarapu et al., 2010]{Akarapu}
Akarapu, S., Sharp, T., Robbins, M.O., 2010. Stiffness of contacts between rough surfaces. arXiv:10.1479v1.

\bibitem[Andersson \& S.~Bj\"{o}rklund, 1994]{bjorklund_andersson1994}
Andersson, S., Bj\"{o}rklund, S., 1994. A numerical method for real elastic contacts subjected to normal and tangential loading. Wear 179, 117--122.

\bibitem[Borri-Brunetto et al., 2001]{Borri}
Borri-Brunetto, M., Chiaia, B., Ciavarella, M., 2001. Incipient sliding of rough surfaces in contact: a multiscale numerical analysis. Comput. Methods Appl. Mech. Eng. 190, 6053--6073.

\bibitem[Bush et al., 1975]{bush1975}
Bush, A.W., Gibson, R.D., Thomas, T.R., 1975. The elastic contact of a rough surface. Wear 35, 87--111.

\bibitem[Bush et al., 1979]{bush1979}
Bush, A.W., Gibson, R.D., Keogh, G.P., 1979. Strongly anisotropic rough surfaces.
Transactions of the ASME. Journal of Lubrication Technology 101, 15--20.

\bibitem[Campa{\~n}\'a \& M\"user, 2006]{campana06}
Campa{\~n}\'a, C., M\"user, M.~H., 2006. Practical Green's function approach to the simulation of elastic semi-infinite solids. Phys. Rev. B 74, 075420.

\bibitem[Campa{\~n}\'a \& M\"user, 2007]{campana07epl}
Campa{\~n}\'a, C., M\"user, M.~H., 2007. Contact mechanics of real vs. randomly rough surfaces: A Green's function molecular dynamics study. Europhys. Lett. 77, 38005.

\bibitem[Campa\~{n}\'{a} et al., 2008]{campanaJPCM08}
Campa\~{n}\'{a}, C., M\"{u}ser, M.H., Robbins, M.O., 2008. Elastic contact between self-affine surfaces: comparison of numerical stress and contact correlation functions with analytic predictions. J. Phys.: Condens. Matter 20, 354013.

\bibitem[Campa\~{n}\'{a} et al., 2011]{Stiff}
Campa\~{n}\'{a}, C., Persson, B.N.J., M\"user, M.H., 2011. Transverse and normal interfacial stiffness of solids with randomly rough surfaces. J. Phys.: Condens. Matter 23, 085001.

\bibitem[Greenwood \& Williamson, 1966]{greenwood1966}
Greenwood, J.A., Williamson, J.B.P., 1966. Contact of nominally flat surfaces. Proc. R. Soc. Lond. A 295, 300--319.

\bibitem[Greenwood \& Tripp, 1970]{greenwood1970}
Greenwood, J.A., Tripp, J.H., 1970. The contact of two nominally flat rough surfaces. Proc. IME 185, 625--633.

\bibitem[Griebel et al., 2007]{Griebel2007}
Griebel, M., Knapek, S., Zumbusch, G., 2007. Numerical Simulation in Molecular Dynamics. Springer, Berlin, Heidelberg.

\bibitem[Hyun et al., 2004]{Hyun} Hyun, S., Pei, L., Molinarie, J.F., Robbins, M.O., 2004. Finite-element analysis of contact between elastic self-affine surfaces. Phys. Rev. E 70, 026117.

\bibitem[Johnson, 1985]{Johnson2}
Johnson, K.L., 1985. Contact Mechanics. Cambridge University Press, Cambridge.

\bibitem[Ju \& Farris, 1996]{ju1996}
Ju, Y., Farris, T.N., 1996. Spectral analysis of two-dimensional contact problems. Trans. ASME. J. Tribol. 118, 320--328.

\bibitem[Kalker, 1977]{kalker1977}
Kalker, J.J., 1977. Variational principles of contact elastostatics. IMA J. Appl. Math. 20, 199--219.

\bibitem[Lorenz \& Persson, 2009]{LP}
Lorenz, B., Persson, B.N.J., 2009. Interfacial separation between elastic solids with randomly rough surfaces: comparison of experiment with theory. J. Phys.: Condens. Matter 21, 015003.

\bibitem[Lorenz et al., 2010]{LCS}
Lorenz, B., Carbone, G., Schulze, C., 2010. Average separation between a rough surface and a rubber block: Comparison between theories and experiments. Wear 268, 984--990.

\bibitem[Lorenz \& Persson, 2010a]{Leak1}
Lorenz, B., Persson, B.N.J., 2010a. Leak rate of seals: Effective-medium theory and comparison with experiment. Eur. Phys. J. E 31, 159--167.

\bibitem[Lorenz \& Persson, 2010b]{Leak2}
Lorenz, B., Persson, B.N.J., 2010b. Europhys. Lett. 90, 38002.

\bibitem[Lorenz \& Persson, 2010c]{squeezeout}
Lorenz, B., Persson, B.N.J., 2010c. Time-dependent fluid squeeze-out between solids with rough surfaces. Eur. Phys. J. E 32, 281--290.

\bibitem[Lubrecht \& Ioannides, 1991]{lubrecht1991}
Lubrecht, A.A., Ioannides, E., 1991. A fast solution of the dry contact problem and the associated sub-surface stress field, using multilevel techniques. Trans. ASME. J. Tribol. 113, 128--133.

\bibitem[Nayak, 1971]{nayak1971}
Nayak, R.P., 1971. Random process model of rough surfaces. Lubr. Technol. Trans. ASME 93, 398--407.

\bibitem[Onions \& Archard, 1973]{onions1973}
Onions, R.A., Archard, J.F., 1973. The contact of surfaces having a random structure.
J. Phys. D 6, 289--304.

\bibitem[Pei et al., 2005]{Pei}
Pei, L., Hyun, S., Molinari, J.F., Robbins, M.O., 2005. Finite element modeling of elasto-plastic contact between rough surfaces. J. Mech. Phys. Solids 53, 2385--2409.

\bibitem[Persson, 2001]{JCPpers} Persson, B.N.J., 2001. Theory of rubber friction and contact mechanics. J. Chem. Phys. 115, 3840;

\bibitem[Persson et al., 2002]{Bucher}
Persson, B.N.J., Bucher, F., Chiaia, B., 2002. Elastic contact between randomly rough surfaces: Comparison of theory with numerical results. Phys. Rev. B 65, 184106.

\bibitem[Persson, 2002]{BNJP}
Persson, B.N.J., 2002. Adhesion between an elastic body and a randomly rough hard surface. Eur. Phys. J. E 8, 385--402.

\bibitem[Persson et al., 2005]{P3}
Persson, B.N.J., Albohr, O., Tartaglino, U., Volokitin, A.I., Tosatti, E., 2005. On the nature of surface roughness with application to contact mechanics, sealing, rubber friction and adhesion. J. Phys. Condens. Matter 17, R1.

\bibitem[Persson, 2006]{PSSR} Persson, B.N.J., 2006. Contact mechanics for randomly rough surfaces. Surf. Sci. Rep. 61, 201.

\bibitem[Persson, 2007]{PerssonPRL}
Persson, B.N.J., 2007. Relation between interfacial separation and load: A general theory of contact mechanics. Phys. Rev. Lett. 99, 125502.

\bibitem[Persson, 2008]{elast}
Persson, B.N.J., 2008. On the elastic energy and stress correlation in the contact between elastic solids with randomly rough surfaces. J. Phys.: Condens. Matter 20, 312001.

\bibitem[Persson \& Yang, 2008]{Chunyan1}
Persson, B.N.J., Yang, C., 2008. Theory of the leak-rate of seals. J. Phys.: Condens. Matter 20, 315011.

\bibitem[Persson, 2010]{Pekl}
Persson, B.N.J., 2010. Fluid dynamics at the interface between contacting elastic solids with randomly rough surfaces. J. Phys.: Condens. Matter 22, 265004.

\bibitem[Poon \& Bhushan, 1995]{poon1995}
Poon, C.Y., Bhushan, B., 1995. Comparison of surface roughness measurements by stylus profiler, {A}{F}{M} and non-contact optical profiler. Wear 190, 76--88.

\bibitem[Rapaport, 2004]{Rapaport2004}
Rapaport, D.C., 2004. The Art of Molecular Dynamics Simulation, second ed. Cambridge University Press, Cambridge.

\bibitem[Ren \& Lee, 1994]{ren1994}
Ren, N., Lee, Si~C., 1994. Effects of surface roughness and topography on the contact behavior of elastic bodies. Trans. ASME. J. Tribol. 116, 804--811.

\bibitem[Sahlin et al., 2010]{sahlin2010a}
Sahlin, F., Larsson, R., Marklund, P., Lugt, P.M., Almqvist, A., 2010. A mixed lubrication model incorporating measured surface topography. Part 1: theory of flow factors. Proc. IME J J. Eng. Tribol. 224, 335--351.

\bibitem[Saito, 2004]{saito04jpsj}
Saito, Y., 2004. Elastic lattice Green's function in three dimensions. J. Phys. Soc. Jap. 73, 1816.

\bibitem[Scaraggi et al., 2011]{mic}
Scaraggi, M., Carbone, G., Persson, B.N.J., Dini, D., 2011. Submitted to Soft Matter.

\bibitem[Stanley \& Kato, 1997]{stanley1997}
Stanley, H.~M., Kato, T., 1997. {F}{F}{T}-based method for rough surface contact. Trans. ASME. J. Tribol. 119, 481--485.

\bibitem[Tian \& Bhushan, 1996]{Tian1996}
Tian, X., Bhushan, B., 1996. A numerical three-dimensional model for the contact of rough surfaces by variational principle. J. Tribol. 118, 33--42.

\bibitem[Westergaard, 1939]{Westergaard_1939}
Westergaard, H.M., 1939. Bearing pressure and cracks. ASME J. Appl. Mech. 6, 49--53.

\bibitem[Whitehouse \& Archard, 1970]{whitehouse1970}
Whitehouse, D.J., Archard, J.F., 1970. The properties of random surface of significance in their contact. Proc. R. Soc. Lond. A 316, 97--121.

\bibitem[Whitehouse \& Phillips, 1978]{whitehouse1978}
Whitehouse, D.J., Phillips, M.J., 1978. Discrete properties of random surfaces. Phil. Trans. R. Soc. Lond. A 290, 267--298.

\bibitem[Whitehouse \& Phillips, 1982]{whitehouse1982}
Whitehouse, D.J., Phillips, M.J., 1982. Two-dimensional discrete properties of random surfaces. Phil. Trans. R. Soc. Lond. A 305, 441--468.

\bibitem[Yang et al., 2006]{Yang2006}
Yang, C., Tartaglino, U., Persson, B.~N.~J., 2006. A multiscale molecular dynamics approach to contact mechanics. Eur. Phys. J. E 19, 47--58.

\bibitem[Yang \& Persson, 2008]{Yang2008}
Yang, C., Persson, B.~N.~J., 2008. Contact mechanics: contact area and interfacial separation from small contact to full contact. J. Phys.: Condens. Matter 20, 215214.
\end{thebibliography}
\end{document}